\documentclass[sigconf]{acmart}

\AtBeginDocument{%
  \providecommand\BibTeX{{%
    \normalfont B\kern-0.5em{\scshape i\kern-0.25em b}\kern-0.8em\TeX}}}

\setcopyright{acmlicensed}
\copyrightyear{2018}
\acmYear{2018}
\acmDOI{XXXXXXX.XXXXXXX}

\acmConference[Conference acronym 'XX]{Make sure to enter the correct
  conference title from your rights confirmation emai}{June 03--05,
  2018}{Woodstock, NY}
\acmISBN{978-1-4503-XXXX-X/18/06}

\usepackage{multirow}
\usepackage{multicol}
\usepackage{float}
\usepackage{microtype}
\usepackage{graphicx}
\usepackage{subfigure}
\usepackage{booktabs} 
\usepackage{enumitem}
\usepackage{wrapfig}

\usepackage{colortbl} 
\usepackage{xcolor} 

\usepackage{algorithm}
\usepackage{algorithmic}
\usepackage{amsmath} 
\usepackage{caption} 
\usepackage{graphicx} 

\theoremstyle{definition}

\theoremstyle{plain}

\begin{document}

\title{Item Cluster-aware Prompt Learning for Session-based Recommendation}

\author{Wooseong Yang}
\affiliation{%
  \institution{University of Illinois Chicago}
  \city{Chicago}
  \country{USA}
}
\email{wyang73@uic.edu}

\author{Chen Wang}
\affiliation{%
  \institution{University of Illinois Chicago}
  \city{Chicago}
  \country{USA}
}
\email{cwang266@uic.edu}

\author{Zihe Song}
\affiliation{%
  \institution{University of Illinois Chicago}
  \city{Chicago}
  \country{USA}
}
\email{zsong29@uic.edu}

\author{Weizhi Zhang}
\affiliation{%
  \institution{University of Illinois Chicago}
  \city{Chicago}
  \country{USA}
}
\email{wzhan42@uic.edu}

\author{Philip S. Yu}
\affiliation{%
  \institution{University of Illinois Chicago}
  \city{Chicago}
  \country{USA}
}
\email{psyu@uic.edu}

\renewcommand{\shortauthors}{Trovato and Tobin, et al.}

\begin{abstract}
    Session-based recommendation (SBR) captures dynamic user preferences by focusing on item sequences within sessions. However, many existing methods primarily model intra-session item relationships (within a single session) while overlooking inter-session relationships (between items across different sessions), limiting their effectiveness to fully capture complex item relations. Although several studies incorporate inter-session information, they often face high computational complexity, leading to prolonged training times and reduced efficiency. To address these issues, we propose the \textbf{CL}uster-aware \textbf{I}tem \textbf{P}rompt learning framework for \textbf{S}ession-\textbf{B}ased \textbf{R}ecommendation (\textbf{CLIP-SBR}) that consists of two modules: 1) \textit{item relationship mining} that incorporates a global graph to extract both intra- and inter-session item relationships effectively from session data, and 2) \textit{item cluster-aware prompt learning} which utilizes soft prompts—lightweight learnable parameters to integrate the mined item relationships into SBR training efficiently. We validate CLIP-SBR through experiments on eight SBR models and three benchmark datasets. Results show consistent improvements in recommendation performance, demonstrating CLIP-SBR’s potential as a robust framework for SBRs. 
\end{abstract}




\received{20 February 2007}
\received[revised]{12 March 2009}
\received[accepted]{5 June 2009}

\maketitle

\newcommand{\checkvalue}[2]{\ifdim #1pt > #2pt \textcolor{blue}{#1}\else \textcolor{red}{#1}\fi}

\section{Introduction} 
Session-based recommendation (SBR) has gained increasing attention due to their effectiveness in various online services~\cite{wang2021survey}, such as e-commerce, social media, and music platforms. Unlike traditional recommendation systems that model long-term static user preferences, SBRs focus on short-term dynamic user preferences embedded within sessions, providing more timely and accurate recommendations. Deep learning-based approaches have recently driven significant advances in SBRs by utilizing their exceptional feature representation capabilities to improve the modeling of complex user preferences embedded in sessions. Specifically, recurrent neural networks (RNNs) have been used to capture sequential dependencies within sessions by processing item sequences in order \cite{hidasi2015session,hidasi2018recurrent}. Several studies employ attention mechanisms to enhance SBRs by selectively focusing on the most relevant items within a session, thereby capturing key patterns in user behavior \cite{li2017neural,hou2022core}. Graph neural networks (GNNs) extend these capabilities by representing sessions as graphs, allowing the models to capture intricate relationships between items through graph structures \cite{wu2019session,yu2020tagnn}.

Despite the progress, most existing SBR approaches primarily focus on the sequential relations between items within a single session (\textit{intra-session} item relationship), often neglecting the valuable information available across multiple sessions (\textit{inter-session} item relationship). This significant \textbf{limitation in session modeling} has been identified and proven in multiple studies \cite{ruocco2017inter,qiu2020exploiting,wang2023modeling,li2023exploiting} as a critical issue in SBRs, hindering the accurate learning of user preferences and potentially affecting recommendation performance. Consider three sessions: session one contains {AirPods Pro, AirPods, AirPods Max}, session two includes {AirPods, Galaxy Buds, Beats Fit Pro}, and session three focuses on unrelated items like running gear and Beats Fit Pro. Intra-session modeling captures sequential patterns but misses broader connections. Inter-session modeling, however, identifies that Beats Fit Pro in session three links to session two, which in turn connects to session one via AirPods. This enables the model to bridge session one and three, revealing deeper item relationships. For instance, the user of session three might also be interested in Airpods Pro instead of Beats Fit Pro, but it’s challenging for the model to identify this solely from intra-session data, as there are no common items between session one and session three. In contrast, inter-session information can discover the deeper connections by linking Beats Fit Pro across session two and three, and AirPods across session one and two, bridging the gap between session one and three.
This highlights the importance of inter-session information, as it enhances the understanding of user preferences beyond what intra-session information alone can achieve. 

Although several SBRs attempt to leverage inter-session relationships, they often face efficiency challenges. For example, GCSAN~\cite{xu2019graph} partially incorporates inter-session information but struggles with limited modeling capabilities, while GCEGNN~\cite{wang2020global} effectively captures complex relationships but incurs high computational costs. These challenges highlight the need for a framework that balances both effectiveness and efficiency.

To address the challenges, we propose \textbf{CL}uster-aware \textbf{I}tem \textbf{P}rompt learning framework for \textbf{S}ession-\textbf{B}ased \textbf{R}ecommendation (\textbf{CLIP-SBR}) that captures complex item relationships and efficiently embeds the information into the SBR training. To overcome the \textbf{limitation in session modeling}, the first module, \textit{item relationship mining}, constructs a global graph from session data. This graph models both intra- and inter-session item relationships in a single structure. Prior studies \cite{wu2019session,xu2019graph,yu2020tagnn,wang2020global,chen2020handling} have demonstrated the effectiveness of graph in capturing item relationships. Inspired by the approaches, the global graph represents items as nodes and item transitions as edges, allowing simultaneous modeling of relationships within and across sessions in a structure. Following graph construction, a community detection method \cite{traag2019louvain} is used to identify item clusters that share similar user preferences. The clusters offer valuable insights, as items within a cluster exhibit close relationships and similar user preferences.

To tackle the \textbf{limitation in efficiency} caused by the challenge of incorporating complex item relationship, we introduce the second module, \textit{item cluster-aware prompt learning}. This module enhances the learning capabilities of SBRs efficiently by integrating learnable soft prompts for each identified item cluster. Prompt learning \cite{liu2023pre}, provides large language models with task-related information through natural language instructions, known as prompts. The concept evolved into learnable prompts, with discrete natural language prompts termed \textit{hard prompt} and continuous, trainable prompts termed \textit{soft prompts}. Soft prompts have demonstrated significant success due to their ability to provide task-related cues to pre-trained models during the tuning phase \cite{li2021prefix, shin2020autoprompt, lester2021power}. This approach has expanded to fields such as computer vision \cite{jia2022visual,zhou2022learning,ge2023domain} and graph representation learning \cite{sun2022gppt,sun2023all,liu2023graphprompt,fang2024universal}, where soft prompts are used to deliver task-specific and/or data-related information during tuning. GPF \cite{fang2024universal} and SUPT \cite{lee2024subgraph} have proven the ability of soft prompt to deliver data-related information during learning process. Inspired by the previous studies, we adopt soft prompts to integrate mined item relationship information into SBR models during training. By tailoring learnable soft prompts to specific item cluster, SBR models can effectively and efficiently learn user preferences from sessions, capturing both intra- and inter-session relationships. 

Unlike previous approaches, our method unifies intra- and inter-session item relationships within a global graph, capturing complex relationships across sessions while preserving local patterns. This enables effective extraction of reusable prior knowledge that enhances SBR training without repeated graph construction or clustering. Furthermore, CLIP-SBR is the first to integrate soft prompts directly into SBR training, leveraging cluster-aware prompts to improve recommendation performance while ensuring adaptability across diverse models, positioning it as a universal framework for session-based recommendation.

To summarize, the main contributions lie in:
 \begin{itemize}[leftmargin=*,  topsep=1pt]
    \item We propose a graph-based method with community detection to design and extract complex item relationships from session data.
    \item We propose a novel approach to enhance SBRs by incorporating cluster-aware item prompt into the model training phase. 
    \item To the best of our knowledge, CLIP-SBR is the first attempt to adopt a soft prompt to the SBR. Also, it is the pioneer work to introduce a soft prompt directly in the model training.
    \item We conduct extensive experiments on eight representative SBRs and three benchmark datasets to demonstrate the superiority of the proposed CLIP-SBR framework.
\end{itemize}


\section{Related Works}
\subsection{Session-based Recommendation}
Session-based Recommendation (SBR) predicts the next user interaction based on recent behavior sequences, effectively capturing short-term user preferences. Recent advancements in deep learning have significantly enhanced SBRs by enabling more sophisticated modeling of item transitions within sessions. GRU4Rec \cite{hidasi2015session} introduced Gated Recurrent Units (GRUs) for sequential dependency modeling, while models like NARM \cite{li2017neural} and STAMP \cite{liu2018stamp} leverage attention mechanisms to capture user intent and preference. CORE \cite{hou2022core} addresses prediction inconsistencies by unifying session and item embeddings, reducing representation gaps. Graph neural networks (GNNs) have further improved SBRs by representing session sequences as graphs, as seen in SRGNN \cite{xu2019graph}, which aggregates item information through gated GNNs, TAGNN \cite{yu2020tagnn} which employs target-aware attention to incorporate user interests. SCL \cite{shi2023self} optimizes the item representation with contrastive learning to improve effectiveness of SBRs. 

While these methods effectively model intra-session relationships, they often overlook the valuable inter-session information, limiting their ability to fully understand complex user behaviors. To address this, several SBRs have been proposed to better model item relationships. GCSAN \cite{xu2019graph} utilizes self-attention on session graphs to capture global intricate item relationships, while GCEGNN \cite{wang2020global} combines item-level and session-level information using graph convolution and self-attention mechanisms. HADCG \cite{su2023enhancing} adopts hyperbolic geometry to extract hierarchical patterns and capture both intra- and inter- session item relationships. Despite their effectiveness, they either struggle to fully capture complex item relationships, which limits their ability to model user preferences and impacts recommendation performance, or suffer from high computational complexity, resulting in prolonged training times. Our work proposes incorporating both intra- and inter-session item relationships using efficient cluster-aware prompts to enhance SBR performance.

\begin{figure*}[t]
\vskip 0.15in
\begin{center}
\centerline{\includegraphics[width=\textwidth]{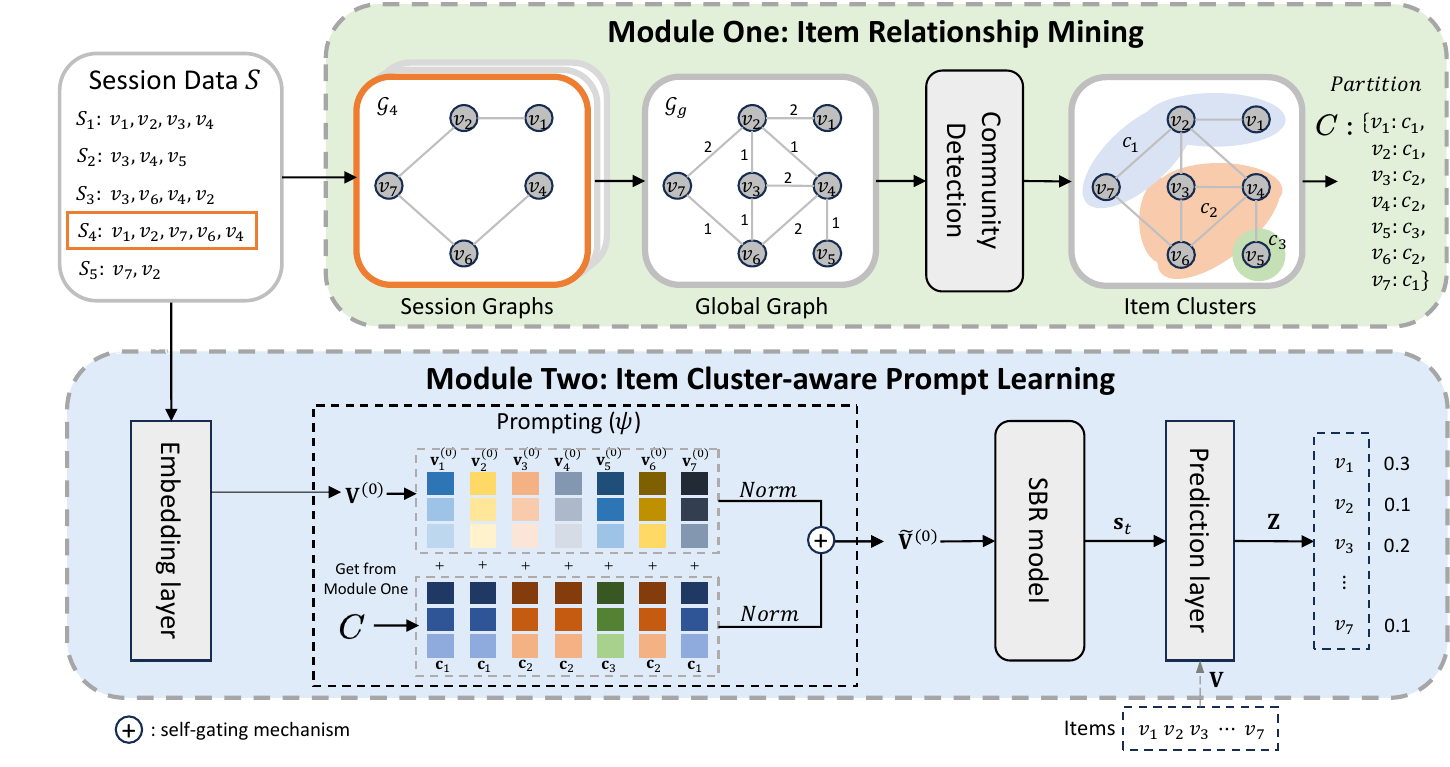}}

\caption{The proposed CLIP-SBR framework consists of two main modules. The \textit{Item Relationship Mining} module constructs session graphs and combines them into a global graph to capture intra- and inter-session item relationships, followed by community detection to identify item clusters. The \textit{Item Cluster-aware Prompt Learning} module enhances SBR models by integrating learnable soft prompts tailored to these clusters, embedding cluster-specific information to improve recommendation accuracy and efficiency.}
\label{fig:framework}
\end{center}
\vskip -0.1in
\end{figure*}

\subsection{Prompt Learning}
Prompt learning is initially introduced in natural language processing (NLP) as a novel method to guide pre-trained language models in performing specific tasks by providing task-related information \cite{liu2023pre}. It originated with the use of discrete hard prompts: text-based instructions designed to provide task related information to pre-trained models \cite{brown2020language}. However, due to the limitations of hard prompts that requires human intervention and limited adaptability, the concept has evolved to include continuous soft prompts. Soft prompts are learnable vectors that diverge from direct natural language representations and are instead optimized during training to enhance model performance on a given task \cite{li2021prefix, shin2020autoprompt, lester2021power}. In recent years, the effectiveness of soft prompts has been demonstrated beyond NLP, extending to domains such as computer vision (CV) and graph representation learning. In the CV field, visual soft prompts have been well-studied and shown great potential in cross-task generalization, domain adaptation, and visual-language models \cite{ge2023domain, zhou2022learning, jia2022visual}. Similarly, in the graph field, pioneering works \cite{sun2022gppt, liu2023graphprompt} align task-specific soft prompts and downstream tasks, such as link prediction. GPF \cite{fang2024universal} and SUPT \cite{lee2024subgraph} further enhance by introducing soft prompts to every node, therby providing graph data-related information in tuning phase, enabling better model adaptation to complex graph structures and tasks. As a result, prompt learning has emerged as a versatile tool that can efficiently improve task performance by providing task- or data-related information across various fields.

\section{Proposed Method}
CLIP-SBR leverages both intra- and inter-session item relationships to effectively capture user preferences embedded in sessions, enhancing the training of existing SBR models with this enriched information. Figure~\ref{fig:framework} presents the overall flow of CLIP-SBR framework, which comprises two main components: 1) \textit{Item Relationship Mining} which constructs a graph from session data to capture item relationships and identifies item clusters using a community detection method. 2) \textit{Item Cluster-aware Prompt Learning} that integrates learnable prompt vectors into item embeddings before they are fed into SBR models. The prompt vectors provide SBR models with information about item clusters during training. In this chapter, we first present the problem statement and then introduce the components of CLIP-SBR in detail.

\subsection{Problem Statement}
Session-based recommendation aims to predict the next item a user will interact with, based on the current session. Let $\mathcal{V} = \{v_1, v_2 \cdots, v_{\mid\mathcal{V}\mid}\}$ represent the set of all unique items across sessions, where $v_k$ represents the $k$-th item ($1 \leq k \leq \mid \mathcal{V} \mid$) in the sequence. We define a session set $\mathcal{S} = \{\mathcal{S}_1,\mathcal{S}_2, \cdots, \mathcal{S}_{\mid 
\mathcal{S} \mid}\}$ and we denote the $t$-th session ($1 \leq t \leq \mid \mathcal{S} \mid$) as $\mathcal{S}_t = \{v^t_1, v^t_2, \cdots, v^t_{\mid \mathcal{S}_t \mid}\}$, where $v^t_l \in \mathcal{V}$ is the $l$-th interacted item ($1 \leq l \leq \mid \mathcal{S}_t \mid$) in $\mathcal{S}_t$. The objective of session-based recommendation is to predict the top-$K$ items ($1 \leq K \leq |V|$) that the user is most likely to interact with next. This translates into forcasting the $({\mid
\mathcal{S}_t\mid} + 1)$-th item based on the first ${\mid
\mathcal{S}_t\mid}$ items in the session.

\subsection{Item Relationship Mining}
As the first module of the CLIP-SBR framework, \textit{Item Relationship Mining} captures complex relationships between items in sessions using a graph-based approach that models both intra- and inter-session item relationships, inspired by prior SBR studies \cite{wu2019session,xu2019graph,yu2020tagnn,wang2020global,chen2020handling}. We first construct session graphs to capture sequential dependencies within individual sessions. To incorporate inter-session relationships, we create a global graph by connecting item nodes if they are linked in any session graph, enabling a unified view of both intra- and inter-session dynamics. By applying a community detection algorithm to this global graph, we identify item clusters—groups of items with strong interconnections and shared user preferences. This clustering step is crucial for uncovering latent patterns and user behaviors, ultimately enhancing the accuracy and efficiency of the SBR training process.

\subsubsection{Graph Construction}
The graph construction process is a foundational step in our framework, designed to model the complex relationships between items in SBRs. We begin by constructing session graphs to capture intra-session item relationships. These session graphs serve as the basis for building a global graph, which captures inter-session item relationships, thereby providing a comprehensive representation of item dependencies across sessions.

\paragraph{Session Graph Construction}
In this step, we transform each session sequence into a session graph, denoted as $\mathcal{G}_t = (\mathcal{V}_t, \mathcal{E}_t)$. Here, $\mathcal{V}_t \subseteq \mathcal{V}$ represents the set of items that a user interacts with within session $\mathcal{S}_t$. The edge set $\mathcal{E}_t = \{ (v^t_i, v^t_j) \mid v^t_i, v^t_j \in \mathcal{V}_t\}$ captures the relationships between these items. Each edge $e^t_{ij}$ is undirected and unweighted, capturing the sequential dependencies between adjacent items $v^t_i$ and $v^t_j$ within the session. This session graph effectively models intra-session item relationships by highlighting the item transitions within each session as edges.

\paragraph{Global Graph Construction}
To capture inter-session item relationships, we construct a global graph by aggregating session graphs with inter-session connections. This global graph, denoted as $\mathcal{G}_g = (\mathcal{V}_g, \mathcal{E}_g)$, includes a node for each unique item in the dataset, where $\mathcal{V}_g = \mathcal{V}$. In this graph, edges are undirected and weighted, with the weight of an edge representing the frequency of item co-occurrences across different sessions. The graph aggregates connections from multiple session graphs, assigning the count of co-occurrences as the weight for each edge. Formally, the edge set is defined as $\mathcal{E}_g = \{(v^g_i, v^g_j) \mid v^g_i, v^g_j \in \mathcal{V}_g\}$, where each edge $e^g_{ij}$ is associated with a weight $w^g_{ij}$, which quantifies the strength of the relationship between items $v^g_i$ and $v^g_j$ based on their co-occurrence frequency. The constructed global graph plays a crucial role in our framework by modeling both intra-session and inter-session item relationships in a single structure. We utilize this graph representation to detect clusters of items that are closely related.

\begin{algorithm}
\caption{The whole procedure of CLIP-SBR}
\label{alg:ip_sbr}

\begin{flushleft}
\textbf{Input:} Sessions $S$, Items $V$ \\
\textbf{Output:} Recommendation lists
\end{flushleft}

\begin{algorithmic}[1]
    \FOR{each session $\mathcal{S}_t \in \mathcal{S}$}
        \STATE Construct session graph $\mathcal{G}_t$ for session $s$;
    \ENDFOR
    \STATE Aggregate session graphs to form global graph $\mathcal{G}_g$;
    \STATE Get $Partition$ by applying $Leiden$ algorithm on $\mathcal{G}_g$ via Eq. (1);
    \FOR{each iteration}
        \STATE Get initial item embeddings $\mathbf{V}^{(0)}$ via Eq. (2);
        \STATE Initialize cluster prompts $\mathbf{C}$ and assign to corresponding items via Eq. (3);
        \STATE Generate prompted item embeddings $\tilde{\mathbf{V}}^{(0)}$, applying normalization, self-gating and prompting function $\psi$ via Eq. (4)-(5);
        \STATE SBR model $f$ takes item set $\mathcal{S}$ and inital item embeddings $\tilde{\mathbf{V}}^{(0)}$, and return session embeddings $\mathbf{S}$ and updated item embeddings $\mathbf{V}$ via Eq. (6);
        \FOR{each item $v_n \in V$}
            \STATE Compute recommendation scores $z_k$ and the probability $\hat{y}_k$ of item $v_k$ via Eq. (7)-(8);
        \ENDFOR
        \STATE Compute loss $\mathcal{L}$ via Eq. (9) and update model parameters
    \ENDFOR
\end{algorithmic}
\end{algorithm}

\subsubsection{Item Cluster Detection}
The process of detecting item clusters within the global graph $\mathcal{G}_g$ plays a crucial role in capturing the complex relationships between items. In this study, a \textit{cluster} refers to a subset of nodes within the graph that are more densely connected to each other than to nodes outside the subset. This high internal connectivity helps uncover latent item dependencies and user behavior patterns, which are essential for modeling complex item relationships embedded in session data. The goal of this step is to identify item clusters that exhibit strong relationships, which are critical for capturing nuanced user preferences in session-based recommendation. To achieve this, we employ the Leiden algorithm \cite{traag2019louvain}, a community detection method known for its superior performance in handling large-scale and complex graphs. The algorithm enhances the Louvain algorithm \cite{blondel2008fast}, by improving detection quality, accelerating convergence speed, and increasing robustness against noise and resolution limits. Furthermore, it effectively handles unconnected and large-scale graphs, both of  which are common in recommendation settings, making the Leiden algorithm particularly suitable for our task of item cluster detection.

We apply the Leiden algorithm to the global graph $\mathcal{G}_g$ to partition items into clusters:
\begin{equation}
    \textit{Partition} = \textit{Leiden}(\mathcal{G}_g) = \{(v^g_i:c_m) \mid v^g_i \in \mathcal{V}_g, c_m \in \mathcal{C}\}.
\end{equation}

Here, $\textit{Leiden}(\mathcal{G}_g)$ represents the execution of the Leiden algorithm on the global graph $\mathcal{G}_g$, yielding a dictionary-like structure \textit{Partition}, in which each item $v^g_i$ is assigned to a specific cluster $c_m$. The set of detected clusters is denoted as $\mathcal{C} = \{c_1, c_2, \dots, c_{\mid \mathcal{C} \mid}\}$, where each $c_m$ represents a distinct cluster, and $\mid \mathcal{C} \mid$ is the total number of clusters. Identifying item clusters enables us to capture intricate item relationships that are overlooked in many previous SBR studies \cite{hidasi2015session,li2017neural,hou2022core,wu2019session,yu2020tagnn}. By leveraging these clusters, we can better tailor the SBR process to reflect underlying user preferences and interactions across sessions. The identified item clusters are subsequently used in the second module of our framework, where they play a crucial role in integrating item cluster-aware prompt learning into session-based recommendation models.

\subsection{Item Cluster-aware Prompt Learning}

In this section, we introduce a method to integrate the mined information about item relationships into SBR models during training. By leveraging learnable prompts, which have proven effective in providing data-related information to models, we enhance the capability of SBRs to capture both intra- and inter-session item relationships.

First, session data is fed into the embedding layer of the existing SBR to obtain the initial item embeddings. Formally, we represent the initial item embeddings as:
\begin{equation}
    \mathbf{V}^{(0)} = \text{EmbeddingLayer}(\mathcal{V}),
\end{equation}
where $\mathbf{V}^{(0)}=\{\mathbf{v}^{(0)}_1, \mathbf{v}^{(0)}_2, \cdots, \mathbf{v}^{(0)}_{\mid \mathcal{V} \mid} \}$ represents the set of initial item embeddings for the items in $\mathcal{V}$. Here, we assign an unique and randomly initialized learnable vector (soft prompt), called \textit{cluster prompt}, to each detected item cluster. The cluster prompt for a cluster $c_m$ is denoted as $\mathbf{c}_m$. For each item $v_k$ in $\mathcal{V}$, we retrieve its corresponding cluster prompt using the partition from the Leiden algorithm:
\begin{equation}
    c_m = \textit{Partition}(v_k),
\end{equation}
where $c_m$ represents the cluster to which item $v_k$ belongs. The set of all cluster prompts is defined as $\mathbf{C} = \{\mathbf{c}_1, \dots, \mathbf{c}_{\mid \mathcal{C} \mid}\}$.

Next, we first normalize the initial item embedding $\mathbf{v}^{(0)}_k$ and the cluster prompt $\mathbf{c}_m$ to ensure balanced contributions. This step prevents the cluster prompt from overwhelming the item embedding and enables the model to learn balanced information. The normalized item embedding $\mathbf{\hat{v}}^{(0)}_k$ and normalized cluster prompt $\mathbf{\hat{c}}_m$ are then combined using a self-gating mechanism:
\begin{equation}
    \psi(\mathbf{\hat{v}}^{(0)}_k) = g_k \cdot \mathbf{\hat{v}}^{(0)}_k + (1 - g_k) \cdot \mathbf{\hat{c}}_m,
\end{equation}
where \( \hat{v}^{(0)}_k = \frac{v^{(0)}_k}{\| v^{(0)}_k \|} \) and \( \hat{c}_m = \frac{c_m}{\| c_m \|} \) represent the normalized versions of the initial item embedding and cluster prompt, respectively, and \( \psi \) denotes the prompting function. The scalar \( g_k \) is a gating variable that controls the contribution of the item embedding and the cluster prompt. This self-gating mechanism dynamically balances the influence of both components based on their relevance, ensuring that the cluster prompt does not dominate and allowing the model to learn effectively.


The set of prompted item embeddings for $\tilde{\mathbf{V}}^{(0)}$ is defined as:
\begin{equation}
\tilde{\mathbf{V}}^{(0)} = \{ \psi(\mathbf{\hat{v}}^{(0)}_1), \psi(\mathbf{\hat{v}}^{(0)}_2), \ldots, \psi(\mathbf{\hat{v}}^{(0)}_{\mid \mathcal{V} \mid}) \}.
\end{equation}

During the test phase, sessions in test dataset may contain unseen items. If a session includes both unseen items and items that are present in the global graph, new edges are added to the global graph to integrate the unseen items. If all items in a session are unseen, we assign the most frequent cluster to these items.

Prompted item embeddings are passed through the existing SBR model during training, providing information about item relationships. Formally, given sessions $\mathcal{S}$ and prompted item embeddings $ \tilde{\mathbf{V}}^{(0)}$, we apply the SBR model $f$ to obtain session embeddings:
\begin{equation}
    \mathbf{S},\mathbf{V} = f(\mathcal{S}, \tilde{\mathbf{V}}^{(0)}),
\end{equation}
where $\mathbf{S}=\{\mathbf{s}_1,\mathbf{s}_2,\cdots,\mathbf{s}_{\mid \mathcal{S} \mid}\}$ is a set of session embeddings with 
 $t$-th element $\mathbf{s}_t$, and $\mathbf{V}=\{\mathbf{v}_1,\mathbf{v}_2,\cdots,\mathbf{v}_{\mid \mathcal{V} \mid}\}$ is a set of updated item embeddings with $k$-th element $\mathbf{v}_k$.

Next, we move to the prediction layer. In this layer, all items in $\mathcal{V}$ are considered candidate items for recommendation. Let $\mathbf{Z}$ represent the set of recommendation scores, where each element $\mathbf{z}_k$ corresponds to the score for the target item $v_k \in \mathcal{V}$. Given session $\mathcal{S}_t$, the score $\mathbf{z}_k$ is computed by taking the inner product between the session embedding and the item embedding:
\begin{equation}
    \mathbf{z}_k = \mathbf{s}_t^\top \mathbf{v}_k,
\end{equation}
where $\mathbf{v}_k$ is the updated item embedding for item $v_k$. The softmax function is then applied to the unnormalized scores $\mathbf{z}_k$ to obtain the final output probabilities:
\begin{equation}
    \hat{y}_k = \text{Softmax}(\mathbf{z}_k),
\end{equation}
where $\hat{y}_k$ denotes the probability of item $v_k$ being the next click in the current session. The loss function used for training is defined as the cross-entropy loss between the predicted probabilities $\hat{y}$ and the one-hot encoded ground truth labels $y$. The cross-entropy loss measures the difference between the true distribution (given by the one-hot encoding) and the predicted distribution (given by the model's output). Formally, the loss is defined as:

\begin{equation}
    \mathcal{L}(y, \hat{y}) = - \sum_{k=1}^{\mid \mathcal{V} \mid} y_k \log (\hat{y}_k),
\end{equation}
where $y_k$ is the one-hot encoded ground truth for item $v_k$ (with $y_k = 1$ for the correct item and $0$ for all others), and $\hat{y}_k$ is the predicted probability that item $v_k$ is the correct next item. 

By incorporating learnable cluster prompts into item embeddings, our approach enhances the SBR model's ability to capture complex item dependencies, resulting in more accurate and effective recommendations.

\begin{table}
\centering
    \caption{Statistics of datasets.}
    \label{tab:stat}
    \begin{tabular}{lllll}
    \hline
    \textbf{Statistic}              & \textbf{Last.fm} & \textbf{Xing}  & \textbf{Reddit} &  \\ \hline
    No. of users           &  966       &  11,399  & 13,850 &  \\
    No. of items           &     38,784    &  59,039  & 21,790 &  \\
    No. of sessions        &    294,402     & 90,286 & 458,292 &  \\
    Avg. of session length &    12.86     &  5.82   & 4.49 &  \\
    Session per user       &    304.76     &  7.92  & 33.09 & \\ \hline
    \end{tabular}
\end{table}
\begin{table*}[ht]
    \centering
    \caption{Comparison of MRR@5 (M@5), Recall@5 (R@5), MRR@10 (M@10), and Recall@10 (R@10) on the three datasets. Our proposed models are highlighted in gray and the best results are highlighted in boldface.}
    \label{tab:overall}
    \setlength{\tabcolsep}{4pt}
    \begin{tabular}{lcccccccccccc}
        \toprule
        \multirow{2}{*}{\textbf{Model}} & \multicolumn{4}{c}{\textbf{Last.fm}} & \multicolumn{4}{c}{\textbf{Xing}} & \multicolumn{4}{c}{\textbf{Reddit}} \\ 
        \cmidrule(lr){2-5} \cmidrule(lr){6-9} \cmidrule(lr){10-13}
        & \textbf{M@5} & \textbf{R@5} & \textbf{M@10} & \textbf{R@10} & \textbf{M@5} & \textbf{R@5} & \textbf{M@10} & \textbf{R@10} & \textbf{M@5} & \textbf{R@5} & \textbf{M@10} & \textbf{R@10} \\ 
        \midrule
        GRU4Rec         & 7.70  & 13.39 & 8.48  & 19.34 & 10.33 & 15.71 & 10.85 & 19.63 & 35.45 & 46.40 & 36.43 & 53.76 \\
        \rowcolor{gray!20}CLIP-GRU4Rec   & 10.07 & 17.24 & 10.94 & 23.87 & 11.12 & 16.90 & 11.69 & 21.19 & 40.21 & 50.47 & 41.06 & 56.85 \\
        \midrule
        NARM            & 5.50  & 9.74  & 6.15  & 14.65 & 9.66  & 15.41 & 10.27 & 20.02 & 33.08 & 44.48 & 34.09 & 52.04 \\
        \rowcolor{gray!20}CLIP-NARM      & 8.10  & 13.96 & 8.88  & 19.87 & 12.04 & 17.81 & 12.31 & 22.08 & 38.50 & 48.79 & 39.38 & 55.39 \\
        \midrule
        CORE            & 10.51 & 17.70 & 11.64 & 25.57 & 10.27 & 17.79 & 11.10 & 24.01 & 29.33 & 46.54 & 30.48 & 55.12 \\
        \rowcolor{gray!20}CLIP-CORE      & 10.67 & 18.11 & 11.58 & 24.99 & 12.65 & 19.59 & 13.35 & 24.86 & 31.33 & 47.56 & 32.37 & 52.05 \\
        \midrule
        SRGNN           & 9.02  & 14.66 & 9.76  & 20.33 & 14.03 & 19.37 & 14.49 & 22.59 & 38.04 & 47.75 & 38.88 & 54.05 \\
        \rowcolor{gray!20}CLIP-SRGNN     & 12.12 & 18.87 & 12.97 & 25.26 & 14.18 & 19.72 & 14.66 & 23.25 & 39.70 & 49.73 & 40.52 & 55.86 \\
        \midrule
        TAGNN           & 9.25  & 15.30 & 10.00 & 21.02 & 12.68 & 18.37 & 13.15 & 21.82 & 40.07 & 51.10 & 40.85 & 56.94 \\
        \rowcolor{gray!20}CLIP-TAGNN     & \textbf{12.82} & \textbf{19.77} & \textbf{13.70} & \textbf{26.50} & 14.70 & 21.07 & 15.25 & 25.18 & \textbf{41.22} & \textbf{52.01} & \textbf{42.02} & \textbf{57.96} \\
        \midrule
        GCSAN           & 4.47  & 8.69  & 5.03  & 12.99 & 13.02 & 16.82 & 13.71   & 24.07  & 36.03 & 44.13 & 36.81 & 50 \\
        \rowcolor{gray!20}CLIP-GCSAN     & 7.05  & 11.59 & 7.68  & 16.36 & \textbf{15.65} & 20.34 & \textbf{16.04} & 23.29 & 39.07 & 48.21 & 39.62 & 52.54 \\
        \midrule
        GCEGNN          & 10.81 & 17.32 & 11.64 & 23.54 & 12.83 & 19.41 & 13.45 & 24.02 & 38.17 & 48.38 & 39.07 & 55.08 \\
        \rowcolor{gray!20}CLIP-GCEGNN    & 11.05 & 17.60 & 11.89 & 23.94 & 14.36 & \textbf{21.58} & 14.96 & \textbf{26.00} & 37.93 & 48.19 & 38.84 & 55.08 \\
        \midrule
        SCL             & 9.99  & 14.90 & 10.49 & 20.31 & 10.58 & 15.88 & 11.23  & 22.84  & 31.25 & 40.43 & 32.66  & 46.54   \\
        \rowcolor{gray!20}CLIP-SCL       & 10.24 & 15.22 & 11.01   & 21.97    & 11.35 & 15.74 & 11.78   & 23.12  & 31.44 & 40.90 & 33.21   & 47.98   \\
        \bottomrule
    \end{tabular}
\end{table*}

\section{Experiments}
We have conducted extensive experiments to evaluate the proposed CLIP-SBR by answering the following four key research questions:
\begin{itemize}[leftmargin=*]
    \item \textbf{RQ1:} Does CLIP-SBR improve the recommendation performance of SBR baseline models?
    \item \textbf{RQ2:} Does the \textit{Item Relationship Mining} module effectively capture and utilize complex item relationships?
    \item \textbf{RQ3:} Does the \textit{Item Cluster-aware Prompt Learning} module effectively and efficiently integrate item relationships into the SBR training?
    \item \textbf{RQ4:} How do different hyper-parameter settings impact the effectiveness of the CLIP-SBR framework?
\end{itemize}

\subsection{Experimental Setup}
\subsubsection{Datasets}
We employ three benchmark datasets that are widely used in the session-based recommendation.
\begin{itemize}[leftmargin=*]
    \item \textit{Last.fm}\footnote{\url{http://ocelma.net/MusicRecommendationDataset/lastfm-1K.html}} contains the complete listening behavior of approximately 1,000 users collected from Last.fm. In this paper, we focus on the music artist recommendation. We consider the top 40,000 most popular artists and group interaction records within an 8-hour window from the same user as a session, following former studies \cite{chen2020handling,guo2019streaming}.
    \item \textit{Xing}\footnote{\url{http://2016.recsyschallenge.com/}} gathers job postings from a social networking platform, including interactions with the postings by 770,000 users. We split user's records into sessions following \cite{quadrana2017personalizing}.
    \item \textit{Reddit}\footnote{\url{https://www.kaggle.com/colemaclean/subreddit-interactions}} is a dataset collected from social media that includes tuples consisting of a user name, a subreddit where the user commented on a thread, and a timestamp of the interaction. The interaction data was segmented into sessions using a 60-minute time threshold, as outlined in \cite{ludewig2019performance}.
\end{itemize}

Using the preprocessed data provided in \cite{pang2022heterogeneous} as a basis, we further process by following the previous studies \cite{chen2020handling,quadrana2017personalizing,wang2020global}, removing sessions with fewer than three interactions to exclude less informative data. Additionally, we retain only users with five or more sessions to ensure sufficient historical data. We allocated 10\% of the sessions as the test set and the penultimate 10\% as the validation set, with the remaining sessions used for training. Furthermore, for a session $\mathcal{S}_t=\{v^t_1, v^t_2, \cdots, v^t_{\mid \mathcal{S}_t \mid}\}$, we generate sequences and corresponding labels by a sequence splitting preprocessing, i.e.,$(\{v^t_1\},v^t_2),(\{v^t_1,v^t_2\},v^t_3),\cdots,(\{v^t_1,v^t_2,\cdots,v^t_{\mid \mathcal{S}_t \mid-1}\},v^t_{\mid \mathcal{S}_t \mid})$ for tra\\ining, valid and test across all the three datasets. The statistics of datasets, after preprocessing, are summarized in Table~\ref{tab:stat}.

Many prior studies use sampled metrics by ranking relevant items alongside a smaller subset of randomly selected items to accelerate calculations. However, this sampling technique can lead to discrepancies compared to the full, unsampled counterparts \cite{krichene2020sampled}. Therefore, in line with previous works \cite{xie2022contrastive}, we assess each method on the entire item set without sampling, ranking all items that the user has not interacted with based on their similarity scores.

\subsubsection{Baselines}
To evaluate the effectiveness of our approach, we compare its performance against a diverse set of well-established methods. The first five baselines focus soley on intra-session information, while the remaining four aim to enhance item relationship by incorporating additional contextual or gloabl information. We assess our method across various model architectures, including those based on RNNs, GNNs, attention mechanisms, and contrastive learning, to demonstrate its broad applicability and effectiveness.
\begin{itemize}[leftmargin=*]
    \item \textbf{GRU4Rec} \cite{hidasi2015session} is a RNN-based model that employs the Gated Recurrent Unit (GRU) to model interaction sequences.
    \item \textbf{NARM} \cite{li2017neural} is a RNN-based model that improves over GRU4Rec by applying an attention mechanism alongside GRU.
    \item \textbf{CORE} \cite{hou2022core} unifies the representation space for both encoding and decoding processes by using a representation-consistent encoder ensuring that session embeddings and item embeddings are in the same space.
    \item \textbf{SRGNN} \cite{wu2019session} transforms session sequences into directed unweighted graphs and applies a gated graph convolutional layer to identify item transition patterns.
    \item \textbf{TAGNN} \cite{yu2020tagnn} is a variant of SRGNN that utilizes target-aware attention network to generate session embeddings by incorporating the features of the candidate item during prediction.
    \item \textbf{GCSAN} \cite{xu2019graph} extracts local context information using a GGNN and then employs a self-attention mechanism to capture explicit dependencies.
    \item \textbf{GCEGNN} \cite{wang2020global} integrates both global context and item sequences from the current session to produce session embeddings through multi-level graph neural networks.
    \item \textbf{SCL} \cite{shi2023self} introduces a simplified contrastive learning framework that enhances item representations without requiring complex sample construction or data augmentation, improving recommendation performance efficiently. In this work, we implemented GCEGNN based SCL for evaluation.
\end{itemize}

\subsubsection{Evaluation Metrics}
To evaluate the performance of the recommendation, we use two metrics widely adopted in SBR evaluation: Mean Reciprocal Rank (MRR@\textit{k}) and Recall (Recall@\textit{k}) following \cite{ruocco2017inter,li2017neural,quadrana2017personalizing}, where $k=5,10$.


\subsubsection{Implementation Details}
We implement CLIP-SBR using a popular open-source recommender systems library \textit{RecBole}\footnote{\url{https://recbole.io}}. All methods are optimized with the Adam optimizer, and we employ early stopping with a patience of 50 epochs to prevent overfitting, using MRR@5 as the indicator. The batch size is tuned from \{64, 128, 256, 512, 1024\} and the learning rate from \{0.001, 0.005, 0.01, 0.05\}. For the Leiden algorithm used in item cluster detection, the resolution value is set to 1, with its impact discussed in detail in Section~\ref{sec:hyper_impact}. To ensure robustness, we run each model five times with different random seeds and report the average performance across these runs.

\subsection{Results}
\subsubsection{Overall Performance (RQ1)}
Table~\ref{tab:overall} presents the experimental results of the eight baselines and their corresponding implementations using the CLIP-SBR framework on three real-world datasets. The best result in each column is highlighted in bold. We applied the CLIP-SBR framework to all baselines, including those that focus solely on intra-session relationships and those that incorporate enhancements to better model complex item relationships to assess its impact. The results show that the CLIP-SBR leads to performance improvements across most of the baselines.

In the intra-session baselines, which focus solely on modeling item relationships within indicidual sessions, showed substantial performance improvements with the application of CLIP-SBR, with the most significant gains observed in the NARM model. Among the inter-session baselines, which incorporate mechanisms to better capture item relationships across sessions, CLIP-SBR demonstrated notable improvements in GCSAN, while having a smaller impact on GCEGNN Further analysis of the impact of applying CLIP-SBR to inter-session baselines is provided in Section~\ref{sec:module2_impact}.

Moreover, the inter-session baselines (GCSAN, GCEGNN) demonstrate generally better performance than the intra-session baselines. Notably, GCEGNN outperforms most intra-session models on the Last.fm dataset. This results highlight the potential of incorporating inter-session information to improve recommendation accuracy, emphasizing the importance of considering both intra-session and inter-session item relationships when developing robust session-based recommendation models.

The application of the CLIP-SBR framework to the intra-session baselines resulted in performance improvements in all case. Specifically, NARM saw a substantial improvement of 45.30\% on Last.fm, 20.11\% on Xing, and 13.04\% on Reddit, marking it as one of the most improved models. TAGNN also demonstrated significant gains, particularly on Last.fm with a 33.91\% increase, 15.31\% on Xing, and 2.33\% on Reddit. Similarly, GRU4Rec achieved notable enhancements of 29.77\% on Last.fm, 7.61\% on Xing, and 11.10\% on Reddit. CORE exhibited improvements of 1.92\%, 16.65\%, and 4.51\% across the respective datasets, while SRGNN showed increases of 31.54\%, 1.44\%, and 4.26\% on the three datasets. SCL showed improvement of 2.33\% on Last.fm, 5.18\% on Xing, and 0.88\% on Reddit. When comparing intra-session CLIP-SBR models with inter-session models, the CLIP-SBR models generally outperformed in many instances. Notably, CLIP-TAGNN consistently surpassed all inter-session models across. CLIP-SRGNN also outperformed inter-session models on Last.fm and Reddit dataset, and outperformed GCSAN in MRR@5 and Recall@5, and GCEGNN in MRR@5. This indicates that the CLIP-SBR framework is highly effective in leveraging and designing inter-session information, leading to better overall performance.

We also applied CLIP-SBR to the inter-session models to assess its effectiveness. The impact varied across the models: it had a substantial effect on GCSAN, and a slight effect on GCEGNN. This suggests that the CLIP-SBR is less beneficial for GCEGNN that already capture inter-session information effectively, while providing significant improvements for GCSAN that do not sufficiently address it. The impact of CLIP-SBR on \textit{inter-session} models is further discussed in Section~\ref{sec:module2_impact}.

\subsubsection{Impact of \textit{Item Relationship Mining} (RQ2)}
We evaluated the impact of the \textit{Item Relationship Mining} module in CLIP-SBR by assessing performance across seven different scenarios.

\begin{figure}[t] 
\centering
\includegraphics[width=\columnwidth]{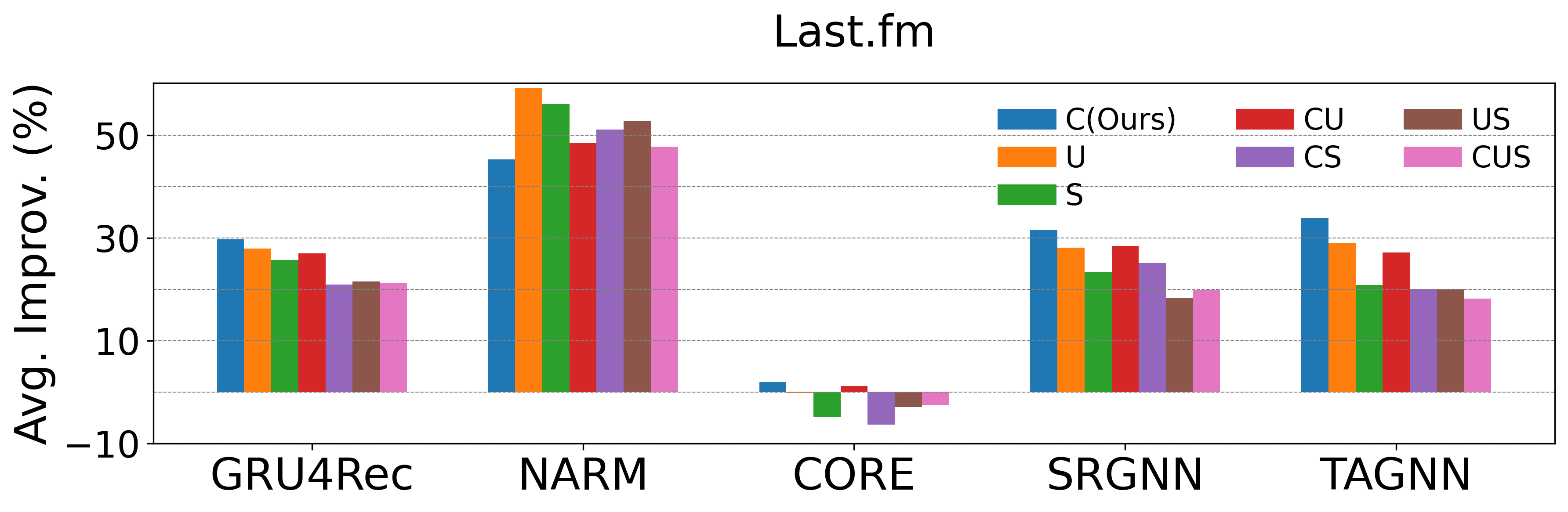}
\includegraphics[width=\columnwidth]{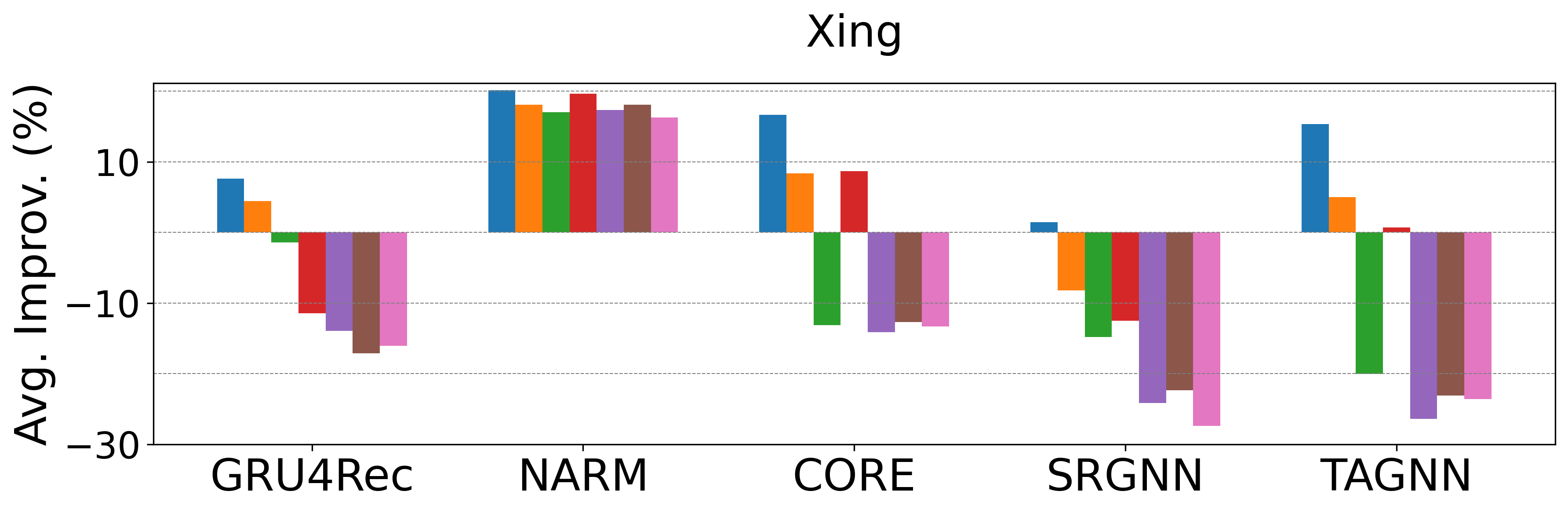}
\includegraphics[width=\columnwidth]{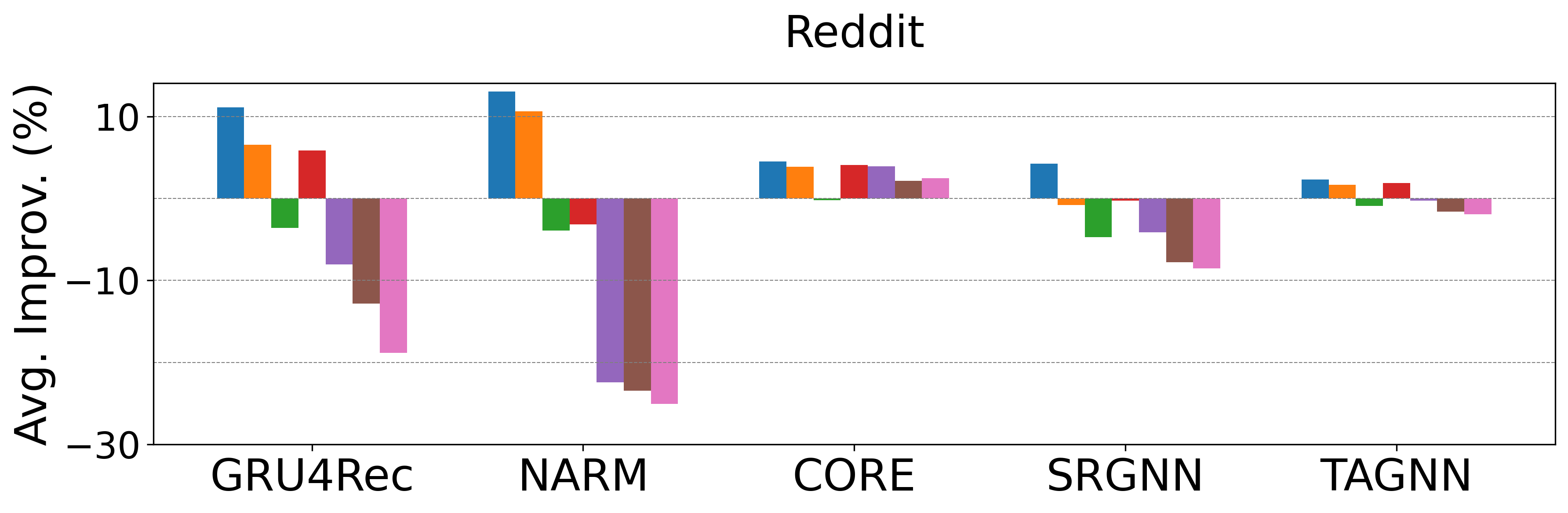}
\caption{
 Improvement (\%) comparison of CLIP-SBR variants on the three datasets. The blue bar represents the original CLIP-SBR, while the other colors indicate various CLIP-SBR variants. "Avg. Improv." refers to the average of improvement percentage calculated from MRR@5 and Recall@5.
}
\label{fig:comparison}
\end{figure}

\begin{itemize}[leftmargin=*]
    \item \textbf{C}: The original CLIP-SBR applies distinct learnable prompts for each item's cluster, embedding them into the item representations. This allows the model to capture group-level item relationships and patterns across sessions.
    \item \textbf{U}: User-specific prompts are added to item embeddings, maintaining a consistent prompt for a given user across all sessions. This helps the model capture user preferences and long-term behavioral patterns across sessions.
    \item \textbf{S}: Session-specific prompts are applied to all items within a session, helping the model learn session-level patterns that reflect user intent or context within a specific session.
    \item \textbf{CU}: Combines both cluster and user-specific prompts. Cluster prompts capture relational structure among items, while user prompts represent long-term user preferences, enabling a deeper understanding of user behavior and item relations.
    \item \textbf{CS}: Combines cluster and session-specific prompts, adding both to item embeddings. This captures the dynamic context of the session (via session prompts) while preserving item relationships within clusters (via cluster prompts).
    \item \textbf{US}: Combines user and session-specific prompts to balance short-term session intent with long-term user preferences. This enables the model to capture how user behavior shifts across sessions while keeping user-specific patterns maintained.
    \item \textbf{CUS}: Combines cluster, user, and session-specific prompts, creating a comprehensive embedding. This setup captures group-level item relationships, user preferences across sessions, and dynamic session behavior simultaneously.
\end{itemize}
We trained these seven types of models for each baselines on each dataset and evaluated their performance improvement using MRR@5 and Recall@5 metrics. The results, as shown in Figure~\ref{fig:comparison}, reveal that in almost all cases, the original CLIP-SBR models with cluster prompts (\textbf{C}) exhibited the most significant improvement compared to the baseline models. Notably, models using user-specific prompts (\textbf{U}) showed the second highest improvement, indicating that incorporating user-specific prompts effectively provided personalizing information during the training phase, thereby enhancing the performance of the SBR models.

In contrast, other scenarios, especially those involving combinations of prompts, did not yield similar improvements and often resulted in performance decreases. This decline can be attributed to the added prompts acting as noise, disrupting the training of the SBR models. For instance, models with combined prompts showed inconsistent or negative performance, as observed in the CORE on the Last.fm dataset and most models on the Xing dataset. This effect is even more pronounced on the Reddit dataset, where most models with combined prompts experienced a decline in performance. 
In summary, among all prompt-based approaches, cluster prompts (\textbf{C}) proved to be the most effective. This outcome suggests that the \textit{Item Relationship Mining} module successfully identifies item communities that share similar user preferences by effectively considering both intra- and inter-session item relationships. These findings highlight the superiority of the module in mining item communities within sessions and underscore its critical role in the CLIP-SBR framework, confirming its effectiveness in enhancing session-based recommendation.

\begin{table}[t]
    \centering
    \caption{Comparison of MRR@5 and average training time (in seconds) for the baselines and proposed methods on the three datasets. Our proposed models are highlighted in gray.}
    \label{tab:training_time}
    \renewcommand{\arraystretch}{0.9} 
    \setlength{\tabcolsep}{2.5pt} 
    \small 
    \begin{tabular}{lcccccc}
        \toprule
        \textbf{Model} & \multicolumn{2}{c}{\textbf{Last.fm}} & \multicolumn{2}{c}{\textbf{Xing}} & \multicolumn{2}{c}{\textbf{Reddit}} \\ 
        \cmidrule(lr){2-3} \cmidrule(lr){4-5} \cmidrule(lr){6-7}
        & \textbf{MRR@5} & \textbf{Time} & \textbf{MRR@5} & \textbf{Time} & \textbf{MRR@5} & \textbf{Time} \\ 
        \midrule
        GRU4Rec & 7.70 & 645.47 & 10.33 & 7.09 & 35.45 & 39.07 \\
        \cellcolor{gray!20}CLIP-GRU4Rec & 10.07 & 745.11 & 11.12 & 6.51 & 40.21 & 30.66 \\ 
        \cmidrule(lr){1-7} 
        NARM & 5.50 & 703.80 & 9.66 & 11.40 & 33.08 & 67.76 \\
        \cellcolor{gray!20}CLIP-NARM & 8.10 & 791.55 & 12.04 & 6.97 & 38.50 & 34.26 \\
        \cmidrule(lr){1-7} 
        CORE & 10.51 & 638.49 & 10.27 & 14.23 & 29.33 & 78.48 \\
        \cellcolor{gray!20}CLIP-CORE & 10.67 & 660.77 & 12.65 & 14.72 & 31.33 & 82.55 \\
        \cmidrule(lr){1-7} 
        SRGNN & 9.02 & 583.72 & 14.03 & 51.94 & 38.04 & 179.29 \\
        \cellcolor{gray!20}CLIP-SRGNN & 12.12 & 619.79 & 14.18 & 56.33 & 39.70 & 196.71 \\
        \cmidrule(lr){1-7} 
        TAGNN & 9.25 & 1227.21 & 12.68 & 174.32 & 40.07 & 250.88 \\
        \cellcolor{gray!20}CLIP-TAGNN & 12.82 & 1416.41 & 14.70 & 197.62 & 41.22 & 290.83 \\ 
        \cmidrule(lr){1-7} 
        SCL & 9.99 & 1042.36 & 10.58 & 152.55 & 31.25 & 202.30 \\
        \cellcolor{gray!20}CLIP-SCL & 10.24 & 1123.44 & 11.35 & 171.20 & 31.44 & 217.65 \\ 
        \midrule[\heavyrulewidth] 
        GCSAN & 4.47 & 1032.88 & 13.02 & 92.70 & 36.03 & 435.70 \\
        \cellcolor{gray!20}CLIP-GCSAN & 7.05 & 1372.21 & 15.65 & 101.75 & 39.07 & 492.67 \\
        \cmidrule(lr){1-7}
        GCEGNN & 10.81 & 625.71 & 12.83 & 60.60 & 38.17 & 260.64 \\
        \cellcolor{gray!20}CLIP-GCEGNN & 11.05 & 645.32 & 14.36 & 63.29 & 37.93 & 269.33 \\
        \bottomrule
    \end{tabular}
\end{table}

\subsubsection{Impact of \textit{Item Cluster-aware Prompt Learning} (RQ3)}
\label{sec:module2_impact}
To evaluate the impact of the \textit{Item Cluster-aware Prompt Learning} module, we assessed its effectiveness on \textit{intra-session} and \textit{inter-session} SBR models. As shown in Table~\ref{tab:overall} and Figure~\ref{fig:comparison}, our approach demonstrates substantial performance improvements, highlighting both the effectiveness and efficiency of the module. In terms of effectiveness, applying the \textit{Item Cluster-aware Prompt Learning} module led to significant performance improvements in \textit{inter-session} baselines. Specifically, CLIP-GCSAN showed notable gains, effectively compensating for GCSAN's limitations in capturing inter-session information, particularly on the Last.fm dataset. For GCEGNN, the module provided moderate improvements, suggesting that while GCEGNN already captures inter-session information effectively, the prompts still offer additional value.

Regarding efficiency, the \textit{Item Cluster-aware Prompt Learning} module demonstrated significant advantages in terms of training time. As shown in Table~\ref{tab:training_time}, several CLIP models achieved better or comparable performance (MRR@5) with significantly less training time compared to the inter-session baselines (GCSAN, GCEGNN). Note that the reported time in Table~\ref{tab:training_time} accounts only for the \textit{Item Cluster-aware Prompt Learning} stage, which is the dominant contributor to training cost; the \textit{Item Relationship Mining} module incurs negligible overhead, taking only about 6 seconds on average.

Among the CLIP models, CLIP-SRGNN shows a good balance between effectiveness and efficiency. It outperformed the inter-session baselines on Last.fm and Reddit, achieving higher MRR@5 with reduced training times—619.79 seconds on Last.fm and 196.71 seconds on Reddit, compared to GCSAN’s 1032.88 seconds, and GCEGNN’s 625.71 seconds on Last.fm, as well as the significantly longer times of the inter-session baselines on Reddit. For the Xing dataset, CLIP-SRGNN achieved comparable performance to the inter-session models while still requiring considerably less time—56.33 seconds versus 92.70 seconds for GCSAN and 60.60 seconds for GCEGNN. The CLIP-TAGNN achieves the best performance on MRR@5 at the expense of longer training time.

This highlights the efficiency of the \textit{Item Cluster-aware Prompt Learning} module, which achieves comparable or superior performance with reduced training time.

\begin{figure}[t] 
\centering
\includegraphics[width=\columnwidth]{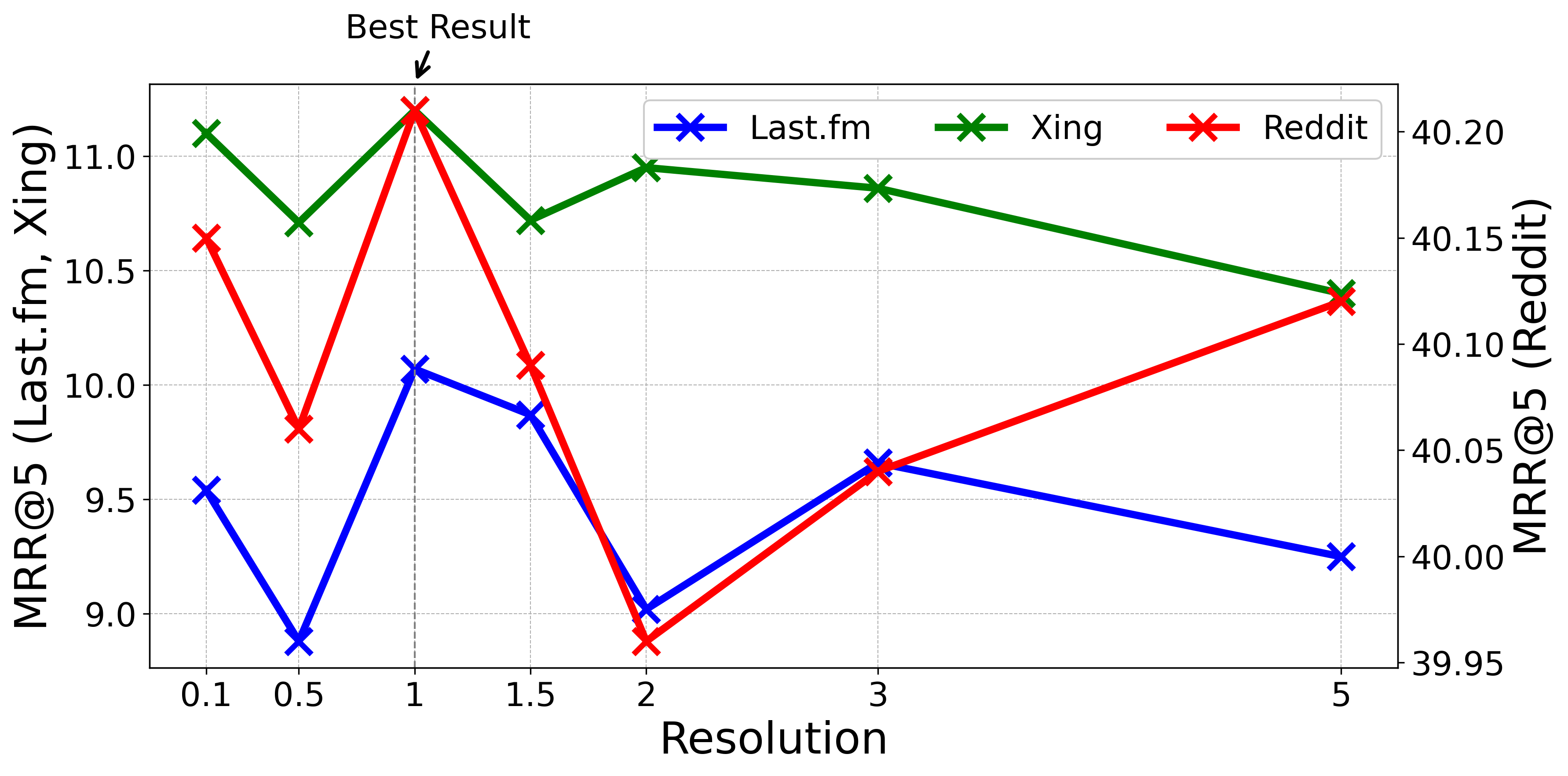}
\caption{
Performance comparison of CLIP-GRU4Rec with different resolution parameters on the three datasets. The MRR@5 scale for Last.fm and Xing is shown on the left y-axis, while the scale for Reddit is on the right y-axis.
}
\label{fig:resolution}
\end{figure}

\subsubsection{Impact of hyperparameter (RQ4)}
\label{sec:hyper_impact} 
The \textit{Item Relationship Mining} module in CLIP-SBR utilizes the Leiden algorithm for item cluster detection, which includes a key hyperparameter called resolution. The resolution parameter controls the granularity of community detection, thereby affecting the size and number of detected communities. To explore how the resolution value influences the performance of CLIP-SBR, we evaluated the performance of CLIP-GRU4Rec with various resolution values: ${0.1, 0.5, 1, 1.5, 2, 3, 5}$. The results, shown in Figure~\ref{fig:resolution}, present the MRR@5 scores across the three datasets. We can observe that while there is no explicit pattern regarding how changes in the resolution value impact performance, the model consistently achieved the best performance when the resolution was set to 1 across all three datasets. This suggests that a resolution value of 1 provides an optimal balance in the granularity of community detection, allowing the model to effectively capture both fine-grained and broader item relationships. 

\section{Conclusion}
In this paper, we introduced the CLuster-aware Item Prompt learning framework for
Session-Based Recommendation (CLIP-SBR), aimed at overcoming the limitations of existing SBR methods by effectively capturing both intra- and inter-session item relationships. The framework comprises two key modules: the first module identifies item clusters from session data, while the second module incorporates cluster-specific prompts to enhance the learning capabilities of SBR models. Extensive experiments demonstrated that CLIP-SBR consistently outperforms baseline models, highlighting its effectiveness. Further analysis of the modules confirmed their significant contributions to the overall performance of CLIP-SBR, with improvements in both recommendation accuracy and computational efficiency. In the future work, we will explore the dynamic prompt learning method that can detect item clusters after each training epoch and assign cluster prompts to each cluster.
\section*{GenAI Usage Disclosure}
During the preparation of this work, generative AI tools were used to support writing-related tasks such as refining sentence structure, improving clarity and coherence, and assisting with LaTeX formatting for tables and equations. All technical contributions, ideas, experiments, and analyses were solely developed and verified by the authors without any assistance from generative AI systems. No generative AI tools were used for coding, data processing, or result generation. All outputs produced with generative AI assistance were carefully reviewed and edited by the authors to ensure accuracy and adherence to academic standards. This use complies with the ACM policy on authorship and the use of generative AI tools in scholarly publishing.


\bibliographystyle{ACM-Reference-Format}
\bibliography{references}

\end{document}